\documentclass[aps, 10pt, twocolumn, superscriptaddress, floatfix]{revtex4-1}
\usepackage{docmute}   
\usepackage{bm, dcolumn, graphicx, amsmath, amssymb, amsfonts, dsfont, kotex, xcolor}

\makeatletter
\renewcommand\paragraph{%
  \@startsection{paragraph}{4}{\parindent}%
    {0.5em plus .05em minus .05em}{-\z@}%
    {\normalfont\normalsize\itshape}%
}
\makeatother
\usepackage[
    colorlinks=true,
    linkcolor=blue,
    citecolor=blue,
    urlcolor=black
]{hyperref}

\usepackage{orcidlink}
\usepackage[capitalize]{cleveref}

\makeatletter
\renewcommand{\bibsection}{%
  \par
  \begingroup
    \baselineskip26\p@
    \bib@device{\columnwidth}{245.5\p@}%
  \endgroup
  \nobreak\@nobreaktrue
  \addvspace{19\p@}%
  \par
}
\makeatother

\newcommand{\burl}[2]{\href{#1}{\textcolor{blue}{#2}}}
\newcommand{\doiref}[2]{%
    {\hypersetup{urlcolor=blue}\href{https://doi.org/#1}{#2}}%
}

\makeatletter
\newcommand{\SaveRevtexFrontmatterCommands}{%
  \global\let\SavedRevtexTitle\title
  \global\let\SavedRevtexAuthor\author
  \global\let\SavedRevtexAffiliation\affiliation
  \global\let\SavedRevtexEmail\email
  \global\let\SavedRevtexThanks\thanks
  \global\let\SavedRevtexDate\date
  \global\let\SavedRevtexMaketitle\maketitle
}
\newcommand{\ResetRevtexFrontmatter}{%
  \frontmatter@init
  \global\let\title\SavedRevtexTitle
  \global\let\author\SavedRevtexAuthor
  \global\let\affiliation\SavedRevtexAffiliation
  \global\let\email\SavedRevtexEmail
  \global\let\thanks\SavedRevtexThanks
  \global\let\date\SavedRevtexDate
  \global\let\maketitle\SavedRevtexMaketitle
}

\newcommand{\SM@prefix}[1]{%
  \def\SM@keys{}%
  \@for\SM@k:=#1\do{%
    \edef\SM@keys{\ifx\SM@keys\@empty\else\SM@keys,\fi SM:\SM@k}}%
}
\newcommand{\SMPrefixCitations}{%
  \let\SM@origcite\cite
  \renewcommand{\cite}[1]{\SM@prefix{##1}\expandafter\SM@origcite\expandafter{\SM@keys}}%
  \let\SM@origbibitem\bibitem
  \renewcommand{\bibitem}[1]{\SM@origbibitem{SM:##1}}%
}
\makeatother

\begin{document}
\SaveRevtexFrontmatterCommands


\title{Interaction-driven ordering of multiple pseudospins and anomalous Hall metals in Bernal-stacked graphene}

\author{Jeonghyeon Suh\,\orcidlink{0000-0001-8377-7402}}
\thanks{These authors contributed equally to this work.}
\affiliation{
Department of Physics and Astronomy,
\burl{https://ror.org/04h9pn542}
{Seoul National University},
Seoul 08826, Korea
}
\affiliation{
Research Institute of Basic Sciences,
\burl{https://ror.org/04h9pn542}
{Seoul National University},
Seoul 08826, Korea
}

\author{Kyungjin Shin\,\orcidlink{0009-0001-2368-0729}}
\thanks{These authors contributed equally to this work.}
\affiliation{
Department of Physics and Astronomy,
\burl{https://ror.org/04h9pn542}
{Seoul National University},
Seoul 08826, Korea
}

\author{Hongki Min\,\orcidlink{0000-0001-5043-2432}}
\email[Contact author: ]{hmin@snu.ac.kr}
\affiliation{
Department of Physics and Astronomy,
\burl{https://ror.org/04h9pn542}
{Seoul National University},
Seoul 08826, Korea
}

\date{\today}

\begin{abstract}

Recent observations of anomalous Hall states in Bernal-stacked tetralayer graphene raise a fundamental question:
How do electron--electron interactions organize multiple low-energy pseudospin sectors within each spin--valley flavor?
Within the Hartree--Fock approximation, we analyze interaction-driven pseudospin ordering and spin--valley polarization as functions of displacement field and carrier density.
We find a rich phase diagram shaped by the interplay between the two pseudospin doublets, including mixed flavor orders at charge neutrality and anomalous Hall metals upon doping.
In particular, we identify topological states with $\sigma_{xy} \approx 6e^{2}/h$, consistent with recent experimental observations.
Our results highlight the central role of multiple pseudospin doublets in the interaction-driven phases of even-layer Bernal multilayers.

\end{abstract}

\maketitle

\paragraph{Introduction}
\label{sec:Intro}---%
Multilayer graphene has emerged as a central platform in condensed matter physics for exploring interaction-driven quantum phases and topology.
The number of layers and the stacking sequence~\cite{Guinea2006,Aoki2007,Min2008a,Min2008b,Koshino2009a,Koshino2009b,Koshino2010,Zhang2010},
as well as the relative twist between adjacent layers~\cite{Bistritzer2011,Tarnopolsky2019,Khalaf2019,Leconte2022,Shin2023,Shin2024,Kim2026,Nakatsuji2023,Devakul2023,Fujimoto2025,Shin2026},
collectively shape its electronic structure.
Combined with electrostatic control of the carrier density and perpendicular displacement field, this structural diversity has enabled the observation of 
superconductivity~\cite{Cao2018,JPark2021,Hao2021,JPark2022,Zhang2022,Zhou2021a,Zhang2023,Yang2025,Patterson2025,Zhang2025,Holleis2025,Han2025,Seo2026},
orbital magnetism~\cite{Lu2019,Polshyn2020,Grover2022,Slot2023,Geisenhof2021,Han2023},
and fractionalization~\cite{Lu2024} in multilayer graphene.

At the microscopic level, many of these quantum phenomena are rooted in the chirality and layer--sublattice character of the low-energy states.
In rhombohedral graphene, these states form a single high-chirality pseudospin doublet with dominant weight on the outermost nondimer orbitals.
Interaction-driven pseudospin order, predicted theoretically~\cite{Min2008c,Zhang2011,Jung2011,MacDonald2012},
has since gained support from a growing body of experiments~\cite{Velasco2012,Shi2020,Han2024a,Liu2024,Sha2024,Han2024b,Dutta2026}.
Stoner-like polarization of the fourfold spin--valley degree of freedom (isospin) can also emerge when a displacement field flattens the low-energy chiral bands~\cite{Yang2025,Zhou2021b,Auerbach2025,Huang2023,Jang2023,Koh2024,Das2024}.

Bernal-stacked graphene, featuring multiple bilayer-like pseudospin doublets in even-layer systems and an additional monolayer-like doublet in odd-layer systems, can host analogous interaction-driven phenomena.
Specifically, Bernal-stacked tetralayer graphene (B4G) has two bilayer-like pseudospin doublets with different effective masses within each spin--valley flavor.
As in bilayer graphene~\cite{Zhou2022,Barrera2022,Seiler2022}, interactions can drive ordered phases in B4G,
as evidenced by anomalous Hall effects~\cite{Chen2023} and a Hall resistance reaching $R_{xy}=\pm h/(6e^2)$ at weak magnetic fields~\cite{Lee2025}.
The structural stability of Bernal stacking also makes these multilayers a favorable platform for exploring interaction physics involving multiple low-energy pseudospin sectors.
However, it remains unclear how the ordering of multiple pseudospin sectors gives rise to these quantum phenomena.

\begin{figure}[t!]
    \includegraphics[width=0.9\linewidth]{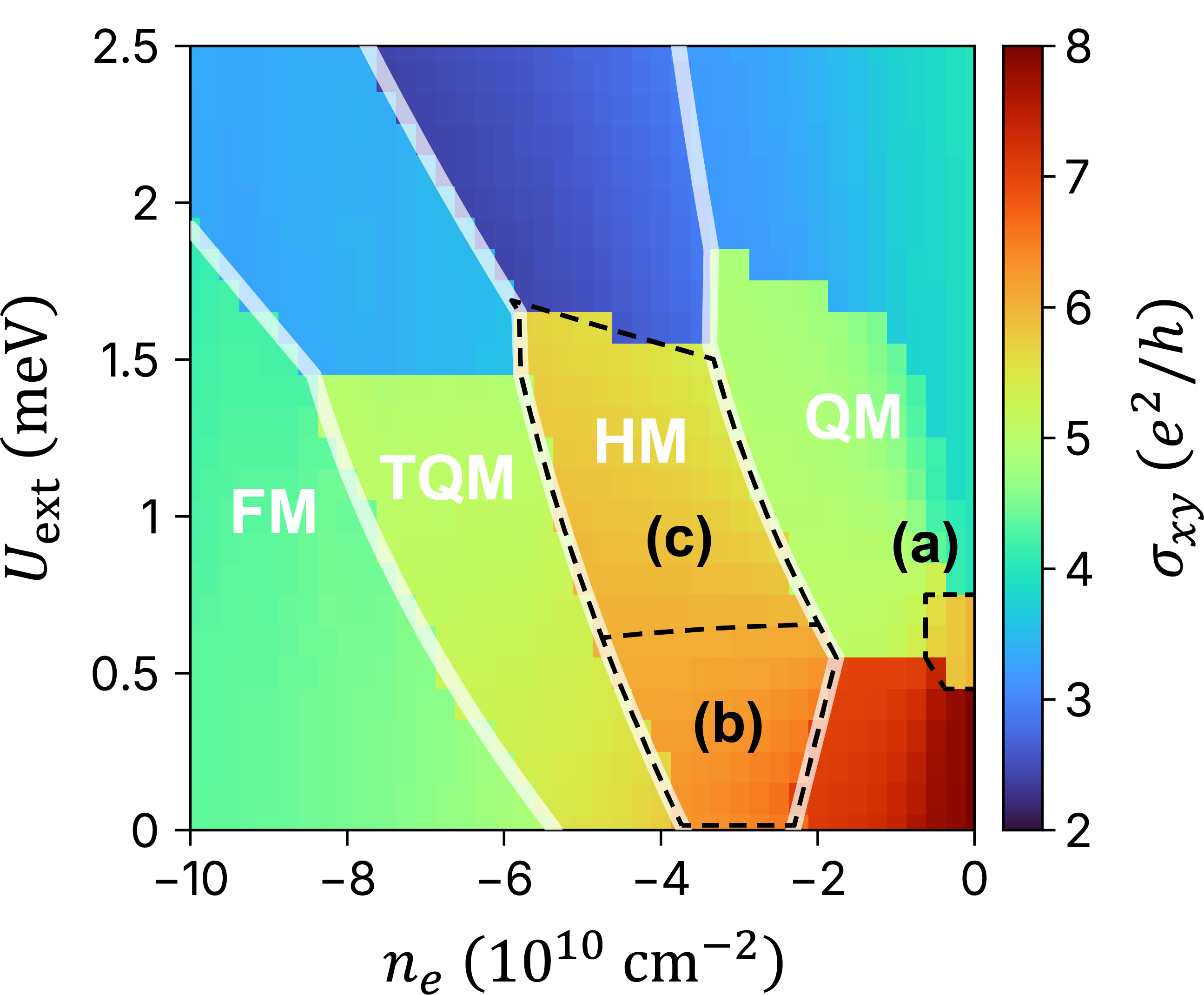}
    \caption{
    Hartree--Fock phase diagram of B4G as a function of the external interlayer potential $U_{\mathrm{ext}}$ and carrier density $n_{e}$ in the hole-doped regime.
    The color scale shows the anomalous Hall conductivity $\sigma_{xy}$ in units of $e^2/h$, assuming the quantum anomalous Hall (QAH) state for the flavor-antiferromagnetic (fAF) order.
    Solid white lines separate quarter-metal (QM), half-metal (HM), three-quarter-metal (TQM), and full-metal (FM) phases.
    Black dashed lines outline three regions with $\sigma_{xy}\approx6e^2/h$:
    (a) the ``fFi+fAF'' mixed order near the charge neutrality point and HM phases within the (b) ``fAF+fAF'' and (c) ``fF+fAF'' orders.
    Orders are listed for the light- then heavy-mass doublets, with fFi (fF) denoting flavor-ferrimagnetic (-ferromagnetic) order.
    Schematics of regions (a)--(c) are shown in Fig.~\ref{fig5}.
    }
    \label{fig1}
\end{figure}

In this paper, we investigate pseudospin ordering and spin--valley polarization in even-layer Bernal multilayers within the Hartree--Fock (HF) approximation, focusing on B4G.
We trace their evolution with displacement field and carrier density, and identify the resulting anomalous Hall states.
At the charge neutrality point (CNP), the two bilayer-like pseudospin doublets form coupled flavor orders and undergo sequential field-induced transitions, leading to mixed-order states.
Upon hole doping, these states evolve through quarter-, half-, three-quarter-, and full-metal phases with progressively weaker spin--valley polarization (see Fig.~\ref{fig1}).
Of particular interest is the Hall response $\sigma_{xy} \approx 6e^{2}/h$, which appears in three regions with distinct microscopic origins:
a mixed pseudospin order near the CNP and two half-metallic phases at finite doping that evolve from different pseudospin orders.
We further show that this picture extends to thicker even-layer Bernal multilayers, providing a common framework for interaction-driven ordering of multiple pseudospin doublets.

\paragraph{Model and methods}
\label{sec:Model}---%
We consider B4G, a four-layer graphene system with ABAB stacking, as illustrated in the insets of Fig.~\ref{fig2}.
\begin{figure}[t!]
    \noindent
    \makebox[\linewidth][l]{\includegraphics[width=0.9\linewidth]{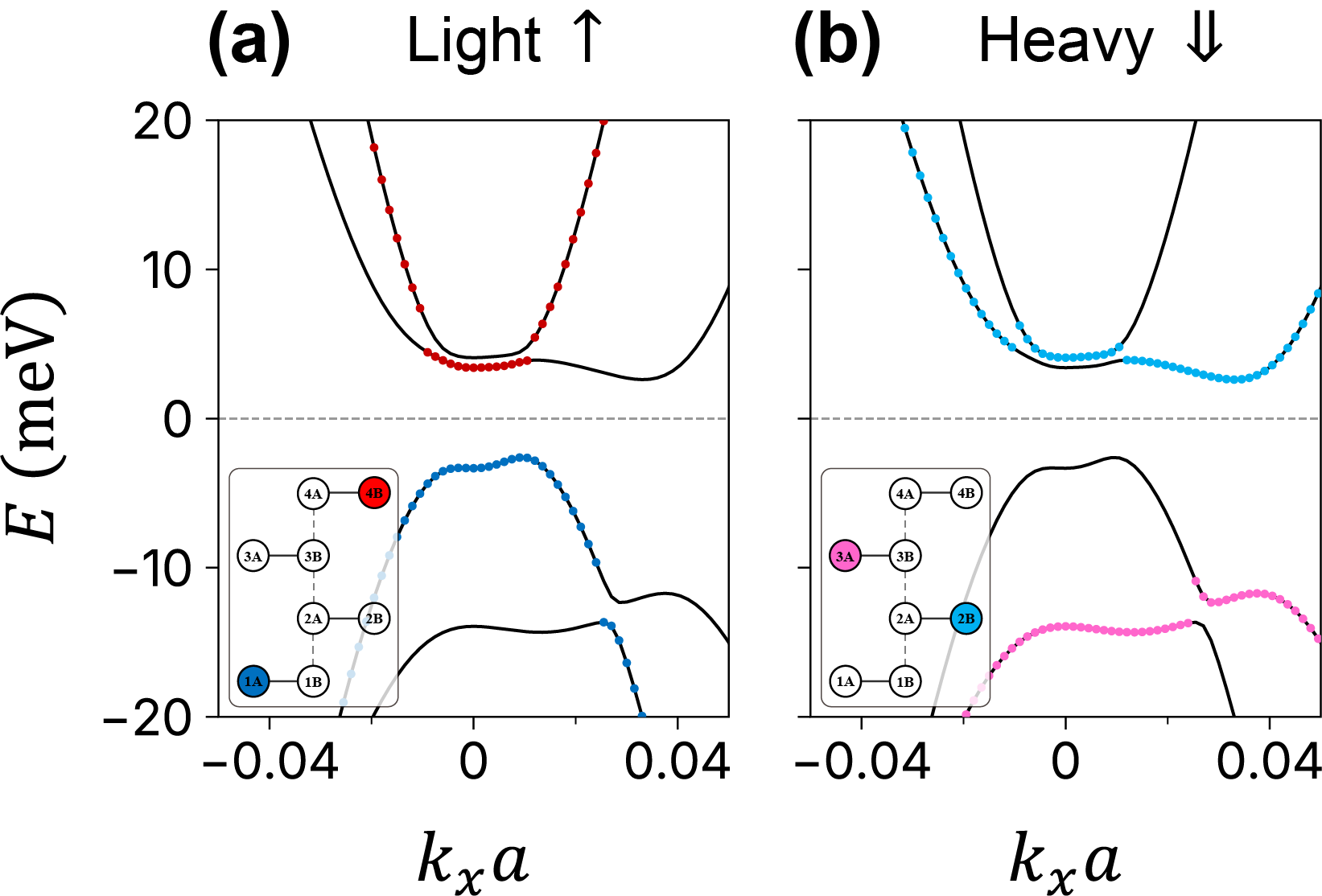}}
    \caption{
    Hartree--Fock band structure of a single spin--valley flavor in the ground state of B4G at the CNP and $U_{\mathrm{ext}}=0$:
    (a) light-mass sector with upward layer polarization ($\uparrow$) and (b) heavy-mass sector with downward layer polarization ($\Downarrow$).
    Bands are colored by their weights on the orbitals shown in the insets, with those of the other sector in black.
    Insets: nondimer orbitals on which each sector is primarily localized, red (pink) on the upper and blue (sky blue) on the lower layer;
    the layer polarization points from the layer hosting the occupied band toward that hosting the unoccupied band.
    }
    \label{fig2}
\end{figure}
The continuum Hamiltonian for the $\pi$ bands is given by
\begin{equation}
    \label{eq:B4G_H0}
    \hat{H}_0
    =
    \sum_{\tau, s, \bm{k}}
    \sum_{\sigma, \sigma^{\prime}}
    \hat{c}_{\tau s \bm{k} ; \sigma}^{\dagger} 
    \varepsilon_{\tau \bm{k} ; \sigma\sigma^\prime}^{(0)} 
    \hat{c}_{\tau s \bm{k} ; \sigma^{\prime}},
\end{equation}
where $\tau=\pm1$ labels the $K$ and $K^\prime$ valleys and $s=\mathrm{u},\mathrm{d}$ labels spin up and down, respectively.
The wavevector $\bm{k}$ is measured from the corresponding valley center,
and $\sigma$ is a composite layer--sublattice index, with the basis ordered as 
$(1\mathrm{A},1\mathrm{B},\ldots,4\mathrm{A},4\mathrm{B})$.
We neglect the weak intrinsic spin--orbit coupling in graphene~\cite{Min2006}, so that $\varepsilon^{(0)}$ is spin independent.
Its explicit form and the Slonczewski--Weiss--McClure parameters are given in the Supplemental Material (SM)~\cite{SM},
with the external potential induced by a perpendicular electric field varying linearly from $-U_{\mathrm{ext}}$ on the bottom layer to $U_{\mathrm{ext}}$ on the top layer.

The low-energy electronic structure of B4G can be understood in terms of two bilayer-like pseudospin doublets with distinct effective masses, both arising from the eigenmodes of the interlayer dimer chain.
The light- and heavy-mass doublets have dominant weight on the outer (1A, 4B) and inner (2B, 3A) nondimer orbitals, respectively.
A perpendicular electric field, entering through $U_{\mathrm{ext}}$, can explicitly break the layer--sublattice symmetry,
whereas electron--electron interactions can break it spontaneously by inducing an imbalance in the occupations of the two orbital components of each pseudospin doublet.
The resulting charge imbalance generates finite layer polarizations of the pseudospins, as illustrated in Fig.~\ref{fig2} by the HF bands obtained below.

We incorporate electron--electron interactions within the HF approximation,
\begin{equation}
    \label{eq:B4G_HMF1}
    \hat{H}_\textrm{MF}
    = \hat{H}_0
    + \sum_{\tau, s, \bm{k}} \sum_{\sigma,\sigma^{\prime}}
    \hat{c}_{\tau s \bm{k};\sigma}^{\dagger}
    \varepsilon_{\tau s \bm{k} ; \sigma\sigma^\prime}^\textrm{(HF)}
    \hat{c}_{\tau s \bm{k} ; \sigma^{\prime}}.
\end{equation}
The matrix elements of the HF term are given by
\begin{align}
    \label{eq:B4G_HMF2}
    \varepsilon_{\tau s \bm{k}; \sigma\sigma'}^\textrm{(HF)}
    & = \delta_{\sigma\sigma^\prime} \sum_{\tau', s', \bm{k}', \sigma''} V_{\ell\ell''}(0) \left< \hat{c}_{\tau' s' \bm{k}' ; \sigma''}^\dagger \hat{c}_{\tau' s' \bm{k}'; \sigma''} \right> \nonumber \\
    & \quad - \sum_{\bm{k}'} V_{\ell\ell'} (|\bm{k} - \bm{k}'|) \left< \hat{c}_{\tau s \bm{k}' ; \sigma'}^\dagger \hat{c}_{\tau s \bm{k}'
    ; \sigma} \right>,
\end{align}
where $\ell$ is the layer index of $\sigma$.
To avoid double counting interaction effects already encoded in the hopping parameters, we employ the charge neutrality scheme~\cite{Huang2024},
in which the density matrix entering Eq.~\eqref{eq:B4G_HMF2} is measured relative to that of the noninteracting ground state at the CNP.
For the electron--electron interaction, we use a layer-dependent Coulomb potential $V_{\ell\ell'}(q)$ screened by symmetric dual gates, as given in the SM.
Intervalley interactions, described by a valley-interchange exchange~\cite{Yang2025,Dutta2026} parameterized as $V_{\mathrm{VI}}=\alpha_{\mathrm{VI}}V_{\ell\ell}(|\bm{K}-\bm{K}'|)$,
with a dimensionless strength $\alpha_{\mathrm{VI}}$, are neglected in the main calculations and examined separately in Fig.~\ref{fig4}.

\paragraph{Pseudospin order at charge neutrality}
\label{sec:Pseudospin}---%
We begin with bilayer graphene, which hosts a single pseudospin doublet.
At zero electric field, the Fock exchange favors spontaneous layer polarization within each spin--valley flavor,
whereas the Hartree interaction penalizes net interlayer charge imbalance.
The resulting ground state is flavor-antiferromagnetic (fAF), with equal numbers of flavors polarized toward opposite layers.
An increasing electric field favors polarization along the field direction, driving successive transitions to flavor-ferrimagnetic (fFi) and flavor-ferromagnetic (fF) states~\cite{Jung2011}.

\begin{figure}[t!]
    \centering
    \includegraphics[width=0.95\linewidth]{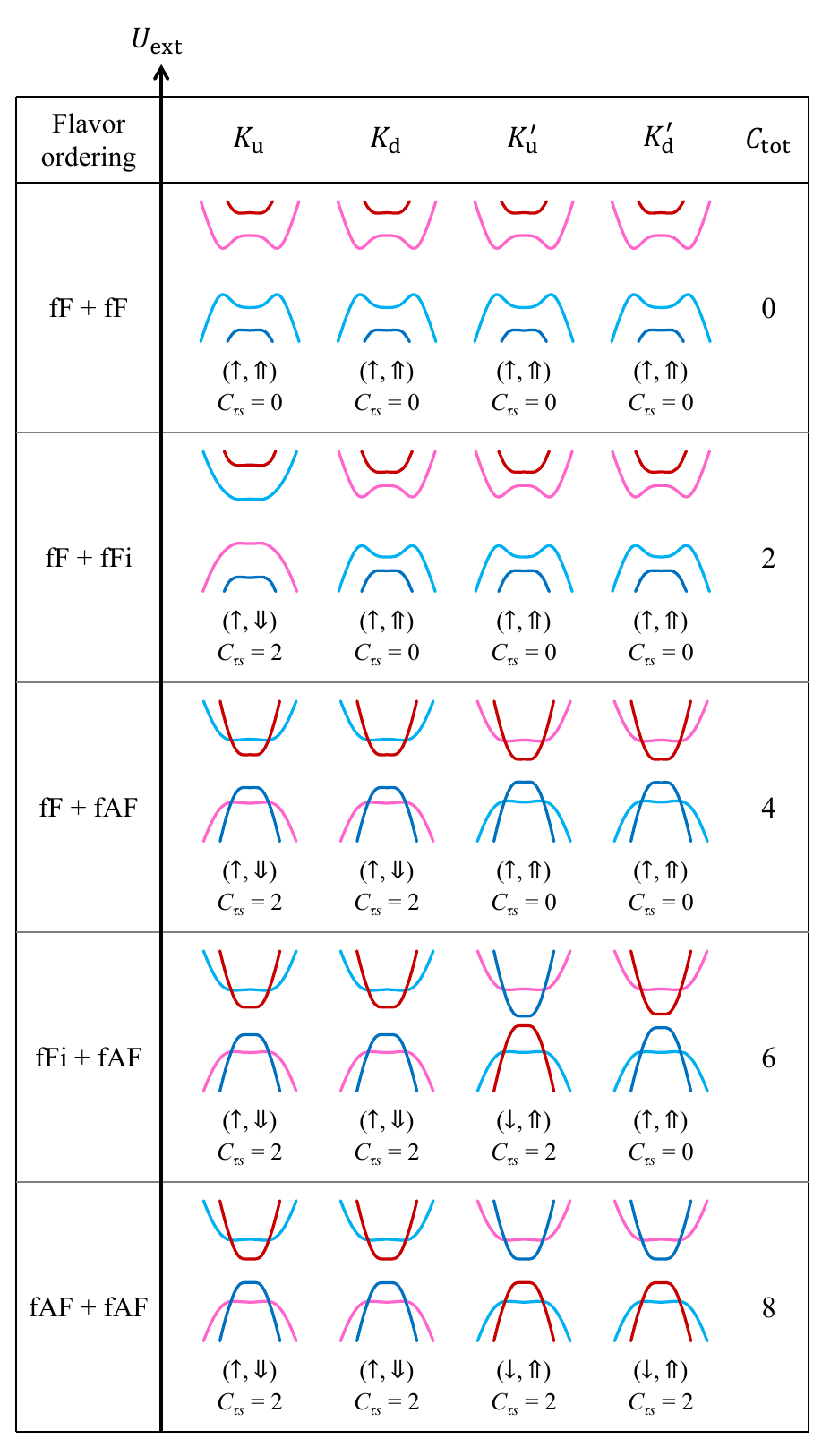}
    \caption{
    Ground-state ordering in B4G at the CNP, with $U_{\mathrm{ext}}$ increasing from bottom to top.
    Horizontal lines separate successive states, with flavor orders listed for the light then heavy sectors.
    Bands are drawn schematically, neglecting remote hopping, with colors and arrow conventions following Fig.~\ref{fig2}.
    Flavor Chern numbers $C_{\tau s}$ are shown below the bands, and their sum $C_{\mathrm{tot}}$ in the last column.
    The sequence starts from the quantum anomalous Hall (QAH) state at $U_{\mathrm{ext}}=0$,
    while interchanging the $K_{\mathrm{d}}$ and $K'_{\mathrm{d}}$ columns gives the sequence starting from the layer-antiferromagnetic (LAF) state,
    with $C_{\mathrm{tot}}=0 \rightarrow 2 \rightarrow 0 \rightarrow 2 \rightarrow 0$ from bottom to top.
    }
    \label{fig3}
\end{figure}

In B4G, this bilayer flavor-ordering picture applies to both the light- and heavy-mass pseudospin doublets.
At zero electric field, both doublets exhibit fAF order, while inter-doublet Fock coupling favors a common sublattice polarization within each spin--valley flavor, corresponding to opposite layer polarizations of the two doublets (see Sec.~S3 of the SM).
With increasing electric field, each doublet undergoes successive fAF--fFi--fF transitions.
Because the light-mass doublet has greater weight on the outer layers, its flavor-order transitions precede those of the heavy-mass doublet, producing mixed-order states with different flavor orders in the two doublets.
The ordering sequence in Fig.~\ref{fig3} follows the effective Hund's rule proposed in Ref.~\cite{Yoon2017}, but includes an intermediate ``fFi+fAF'' state without reentrant ordering.
Trigonal warping has been proposed to yield $C_{\mathrm{tot}}=6$ through asynchronous mass inversions at the central and satellite Dirac cones~\cite{Feng2026}.
However, for the parameters considered, our calculations instead find a transition to the ``fF+fFi'' state before the central-cone mass of the heavy sector inverts.
Details of the ordering sequence and the energetics of competing states are presented in Sec.~S2 of the SM.

We emphasize that the same ordering principle extends to thicker even-layer Bernal multilayers.
In the absence of an electric field, all bilayer-like doublets exhibit fAF order, and within each spin--valley flavor, inter-doublet exchange makes their layer polarizations alternate in order of effective mass.
As the field increases, flavor-order transitions begin in the lightest doublet and proceed to heavier ones.
Intermediate-mass doublets, however, have smaller dipole moments than the lightest doublet and smaller exchange-induced gaps than the heaviest doublet, gaining relatively little energy from either field alignment or exchange.
They tend to reorder together with the heaviest doublet, skipping some of their flavor orders.
The ground-state ordering sequence is therefore dominated by transitions in the lightest and heaviest doublets.

\paragraph{Spin--valley polarization}
\label{sec:Isospin}---%
Exchange interactions favor occupying fewer spin--valley flavors, stabilizing polarized phases at low carrier density when the Fock-energy gain outweighs the kinetic-energy cost.
As the density increases, filling more dispersive parts of the bands raises the kinetic-energy cost of polarization,
driving sequential transitions from a quarter metal (QM) to a half metal (HM), a three-quarter metal (TQM), and finally an unpolarized full metal (FM).
A larger pseudospin gap flattens the band edges and lowers this cost, extending the polarized phases to higher densities.
Chen \textit{et al.}\ reported evidence for a valley-polarized HM in B4G at finite displacement fields~\cite{Chen2023},
and related isospin-polarized phases in rhombohedral graphene have attracted considerable attention~\cite{Yang2025,Zhou2021b,Auerbach2025,Huang2023,Jang2023,Koh2024,Das2024}.

These spin--valley polarized phases indeed emerge as ground states of B4G over a broad region of the HF phase diagram shown in Fig.~\ref{fig1}.
At finite $U_{\mathrm{ext}}$, the fAF order splits the four spin--valley flavors into two pairs,
$(\downarrow,\Uparrow)$ and $(\uparrow,\Downarrow)$ in the notation of Fig.~\ref{fig3}, whose light-sector gaps are reduced and enhanced by the field, respectively.
Holes therefore populate the $(\downarrow,\Uparrow)$ pair first, and this splitting determines which flavors host them along the QM--HM--TQM--FM sequence.
Where these gaps grow with $U_{\mathrm{ext}}$, the phase boundaries accordingly shift to higher hole densities.
Along the QM--HM boundary, however, the light-sector polarizations of the $(\downarrow,\Uparrow)$ pair reverse through the ``fFi+fAF'' state to form $(\uparrow,\Uparrow)$,
so that their gaps first decrease and then increase, producing a kink near $U_{\mathrm{ext}}\approx0.5$~meV.
The valence-band edges of the $(\uparrow,\Uparrow)$ pair shift downward faster with $U_{\mathrm{ext}}$ owing to their larger layer dipoles, becoming comparable to those of the $(\uparrow,\Downarrow)$ pair near $U_{\mathrm{ext}}\approx1.5$~meV.
Beyond this point, the flavors hosting the holes in the QM, HM, and TQM phases switch to the $(\uparrow,\Downarrow)$ pair, producing the jumps in $\sigma_{xy}$.

\paragraph{Anomalous Hall responses}
\label{sec:QAH}---%
The fAF order remains degenerate with respect to how the four spin--valley flavors are assigned to the two pseudospin orientations.
Two representatives are the layer-antiferromagnetic (LAF) state, in which flavors with the same spin are aligned,
and the quantum anomalous Hall (QAH) state, in which flavors in the same valley are aligned.
Intervalley interactions and an external magnetic field lift this degeneracy, favoring the LAF and QAH states, respectively.

\begin{figure}[t!]
    \hspace*{-0.8cm}%
    \includegraphics[width=0.8\linewidth]{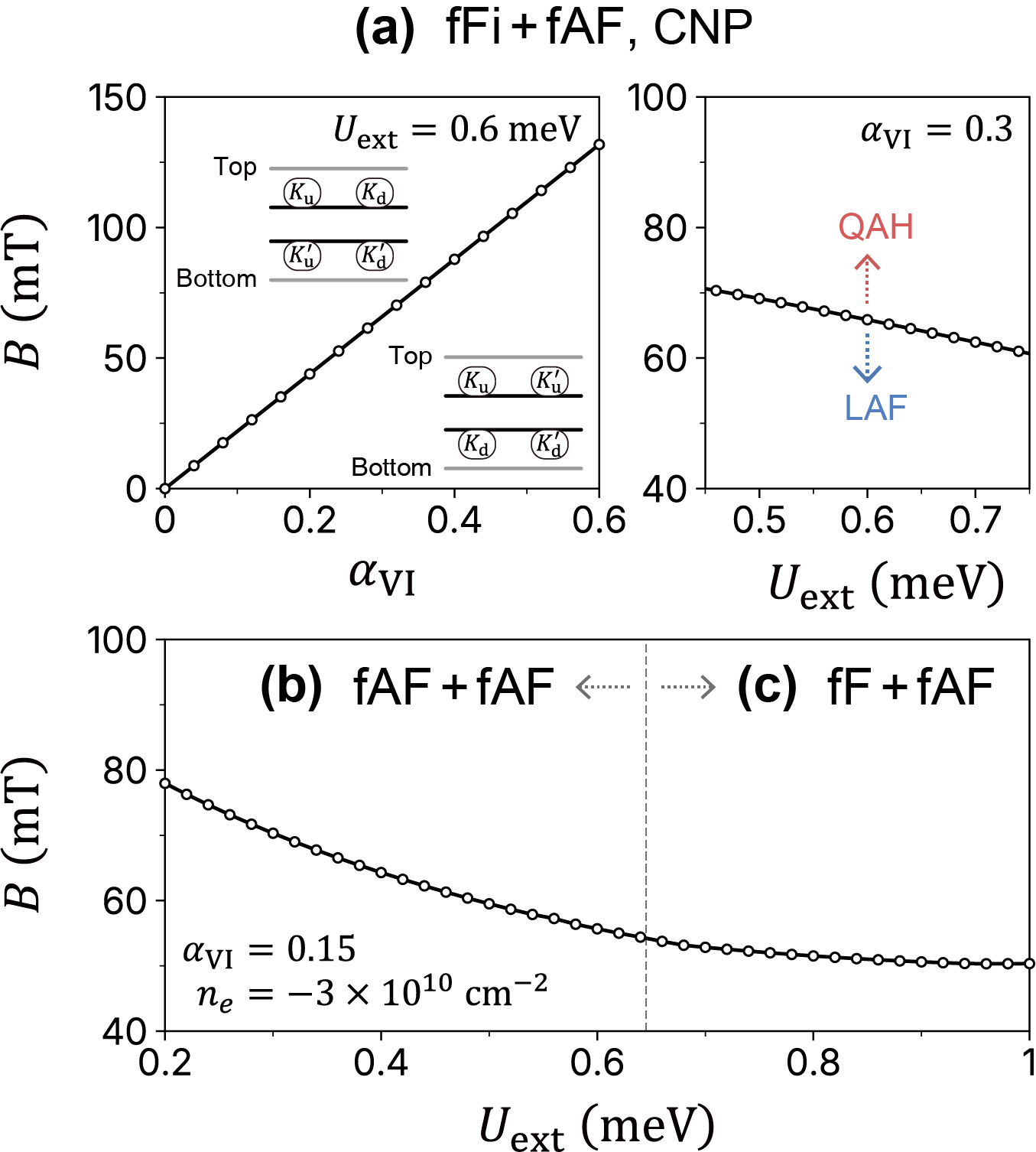}
    \caption{
    Critical field $B_{\mathrm{c}}$ of the LAF--QAH competition in regions (a)--(c) of Fig.~\ref{fig1}; the QAH (LAF) state is favored above (below) each line.
    (a) ``fFi+fAF'' order at the CNP: 
    $B_{\mathrm{c}}$ versus $\alpha_{\mathrm{VI}}$ at $U_{\mathrm{ext}}=0.6$~meV (left) and versus $U_{\mathrm{ext}}$ at $\alpha_{\mathrm{VI}}=0.3$ (right), within the range where this order is the ground state at $\alpha_{\mathrm{VI}}=B=0$.
    Left insets: heavy-sector QAH (upper) and LAF (lower) configurations, with each flavor on the inner layer hosting its occupied band.
    (b),(c) Valley-polarized HMs in the ``fAF+fAF'' and ``fF+fAF'' orders, respectively, at $\alpha_{\mathrm{VI}}=0.15$ and $n_e=-3\times10^{10}$~cm$^{-2}$, separated by the dashed line.
    Values of $\alpha_{\mathrm{VI}}$ are chosen to keep $B_{\mathrm{c}}$ below $0.1$~T, as in experiment.
    }
    \label{fig4}
\end{figure}

Figure~\ref{fig4} quantifies this competition in the three regions of Fig.~\ref{fig1}.
In the weak-field approximation, the QAH state is stabilized above $B_{\mathrm{c}}=\Delta \mathcal{E}/|\Delta \mathcal{M}|$,
where $\Delta \mathcal{E} \propto V_{\mathrm{VI}}$ and $\Delta \mathcal{M}$ are the QAH--LAF differences in HF energy and orbital magnetization, respectively.
The orbital magnetization is evaluated as in Ref.~\cite{Ceresoli2006}, with the chemical potential at midgap for the insulating state.
For the ``fFi+fAF'' order at the CNP [region (a)], whose QAH state carries $C_{\mathrm{tot}}=6$,
$B_{\mathrm{c}}$ grows linearly with $\alpha_{\mathrm{VI}}$ and remains below $0.1$~T up to $\alpha_{\mathrm{VI}}\simeq0.45$ at $U_{\mathrm{ext}}=0.6$~meV, i.e., for $V_{\mathrm{VI}}$ up to about half the Coulomb interaction at $q=|\bm{K}-\bm{K}'|$.
At fixed $\alpha_{\mathrm{VI}}$, $B_{\mathrm{c}}$ decreases with increasing $U_{\mathrm{ext}}$ [Fig.~\ref{fig4}(a)].
The hole-doped half metals in regions (b) and (c) show the same trend, $B_{\mathrm{c}}$ decreasing monotonically as $U_{\mathrm{ext}}$ increases, more gradually in region (c) [Figs.~\ref{fig4}(b) and \ref{fig4}(c)].

This trend reflects $\Delta\mathcal{M}$, as $\Delta\mathcal{E}$ varies only weakly with $U_{\mathrm{ext}}$.
In regions (a) and (b), the orbital magnetization grows as the light-mass gap narrows, whereas in region (c), its reduction from the widening gap is slightly outweighed by the contribution of the hole pocket, whose Berry curvature is reversed at the gap closing.
The decrease is $B_{\mathrm{c}}$ not generic: in the ``fF+fAF'' order at the CNP, for example, $B_{\mathrm{c}}$ instead increases with $U_{\mathrm{ext}}$, as the widening light-mass gap reduces the orbital magnetization without such compensation.
In all three regions relevant to experiment, however, $B_{\mathrm{c}}$ decreases with increasing displacement field, consistent with the observed shift of Hall quantization below $0.1$~T to lower magnetic fields~\cite{Lee2025}.
We henceforth assume the QAH configuration in the Hall calculations, without including intervalley interactions or magnetic fields explicitly.

\begin{figure}[t!]
    \centering
    \includegraphics[width=0.9\linewidth]{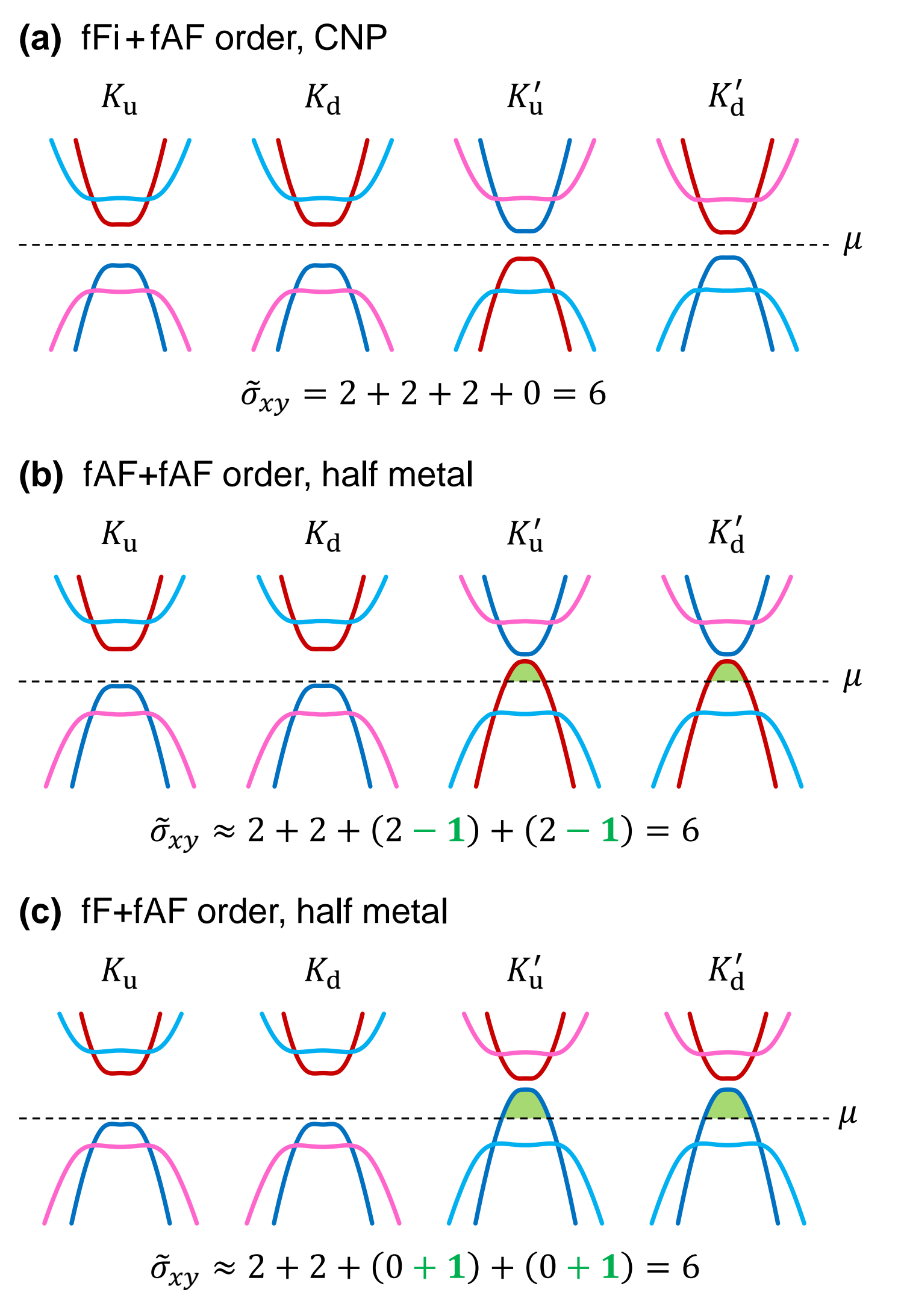}
    \caption{
    Schematics of regions (a)--(c) of Fig.~\ref{fig1}.
    Panels (b) and (c) depict valley-polarized HMs;
    for $U_{\mathrm{ext}}\lesssim0.1$~meV, the ground state in (b) is instead spin-polarized, with the same $\tilde{\sigma}_{xy}$.
    Band colors follow Fig.~\ref{fig2}, and dashed lines mark the chemical potential $\mu$.
    Below each panel, $\tilde{\sigma}_{xy}$ is decomposed into flavor contributions listed from left to right.
    Green shading and numbers indicate hole pockets and the changes in $\tilde{\sigma}_{xy}$ due to hole doping, respectively.
    }
    \label{fig5}
\end{figure}

In the QAH state, broken time-reversal symmetry allows the Hall contributions from the two valleys to add rather than cancel, yielding an anomalous Hall response.
The anomalous Hall conductivity normalized by $e^{2}/h$ is given by
\begin{equation}
    \tilde{\sigma}_{xy} \equiv \frac{\sigma_{xy}}{e^2 / h} = \frac{1}{2\pi} \sum_{\tau, s, n} \int d^2 k \; f_{\tau s\bm{k} ; n} \Omega_{\tau s\bm{k} ; n}^{(z)},
\end{equation}
where $f_{\tau s\bm{k} ; n}$ and $\Omega_{\tau s\bm{k} ; n}^{(z)}$ denote the Fermi--Dirac distribution and the $z$-component of the Berry curvature, respectively, for band $n$.
For an insulating state at the CNP, $\tilde{\sigma}_{xy}$ reduces to the integer Chern number, whereas carrier doping drives it away from this quantized value.

The color scale in Fig.~\ref{fig1} shows broad Hall-response regions delineated by the phase boundaries, reflecting the spin--valley polarization of the underlying states.
Within these regions, hole doping depopulates states near the gapped Dirac points, shifting $\tilde{\sigma}_{xy}$ by approximately $\pm1$ for each spin--valley flavor involved.
The phase diagram offers three scenarios for the experimentally relevant value $\tilde{\sigma}_{xy}\approx 6$~\cite{Lee2025} [regions (a)--(c) of Fig.~\ref{fig1}; schematics in Fig.~\ref{fig5}].
The ``fFi+fAF'' order at the CNP directly yields $\tilde{\sigma}_{xy}=6$ [region (a)].
The other two are realized in the hole-doped half-metallic phases.
At low displacement field, hole doping in the ``fAF+fAF'' order depopulates two light-mass sectors with Chern number $+1$ each, reducing $\tilde{\sigma}_{xy}$ to $\approx8-2=6$ [region (b)].
At high field, the same depopulation in the ``fF+fAF'' order removes $-1$ each, raising it to $\approx4-(-2)=6$ [region (c)].

\paragraph{Discussion}
\label{sec:Discussion}---%
Once electron--electron interactions spontaneously break sublattice symmetry, 
nontrivial band topology arises from the intrinsic valley chirality of multilayer graphene, with the net Chern number determined by spin--valley flavor ordering.
For rhombohedral $N$-layers, a single chirality-$N$ pseudospin doublet undergoes the fAF--fFi--fF sequence,
accompanied by the quantized Hall sequence $\tilde{\sigma}_{xy}=2N\rightarrow N\rightarrow0$ when starting from the QAH state.
Bernal-stacked graphene with an even number of layers departs from this simple picture by hosting multiple bilayer-like pseudospin doublets within each flavor.
Inter-doublet exchange locks their sublattice polarizations so that the Chern contributions add rather than cancel.
Because of their distinct layer profiles, these doublets respond differently to a displacement field,
producing mixed-order states and a richer Hall sequence whose intermediate values are selected by the relative energies of the competing ordered states.
Upon doping, these states evolve into spin--valley polarized anomalous Hall metals with diverse Hall responses.

The multi-doublet picture of pseudospin ordering holds in the interaction-dominant regime, where the light-mass band edges lie well inside those of the heavy-mass bands.
The remote hopping $\gamma_{2}$ counteracts this hierarchy by shifting the heavy-mass bands downward relative to the light-mass ones and mixing the two doublets.
The hierarchy is thus lost as the interaction weakens or upon electron doping.
We focus on the hole-doped side, where the light- and heavy-mass valence-band edges are further separated,
limiting inter-doublet mixing and keeping this side in the interaction-dominant regime for the parameters considered.
Several studies note that $|\gamma_2|$ may be much smaller than previously estimated~\cite{Nam2016,Nam2018} and can be further suppressed by hBN encapsulation, as reported in ABA trilayer graphene~\cite{Che2020}.
The electron-doped side could then also lie within this regime, yielding a phase diagram qualitatively similar to that on the hole-doped side.
Thicker even-layer multilayers follow the same picture, although its range of validity narrows as the interaction-induced gaps decrease with layer number (see Fig.~S1(a) of the SM~\cite{SM}).

In summary, our HF calculations reveal how electron--electron interactions generate anomalous Hall states in Bernal-stacked graphene through the coupled ordering of multiple pseudospin doublets.
The role of orbital magnetic moments is manifest in the weak-field stabilization of quantum anomalous Hall order.
In tetralayer graphene, the experimentally relevant Hall response $\sigma_{xy}\approx6e^{2}/h$ points to three possible scenarios: a mixed pseudospin order near the CNP or one of two distinct half metals at finite doping.
The multi-doublet structure of Bernal graphene ties interaction-driven pseudospin order to gate-tunable anomalous Hall responses.

\medskip
\paragraph{Acknowledgments}
---%
This work was supported by the National Research Foundation of Korea (NRF) through Grant No.~RS-2023-NR076715,
by the Creative-Pioneering Researchers Program through Seoul National University (SNU),
and by the Center for Theoretical Physics.

\medskip
\paragraph{Data availability}
---%
The data that support the findings of this study are available upon reasonable request from the authors.

\twocolumngrid

\onecolumngrid  
\clearpage
\ResetRevtexFrontmatter
\onecolumngrid

\setcounter{section}{0}
\setcounter{equation}{0}
\setcounter{figure}{0}
\setcounter{table}{0}
\renewcommand{\thesection}{S\arabic{section}}
\renewcommand{\theequation}{S\arabic{equation}}
\renewcommand{\thefigure}{S\arabic{figure}}
\renewcommand{\thetable}{S\arabic{table}}
\renewcommand{\theHsection}{S\arabic{section}}
\renewcommand{\theHequation}{S\arabic{equation}}
\renewcommand{\theHfigure}{S\arabic{figure}}
\renewcommand{\theHtable}{S\arabic{table}}

\makeatletter
\def\combined@first{FirstPage}
\def\combined@last{LastBibItem}
\let\combined@origlabel\label
\def\label#1{%
  \def\combined@tempa{#1}%
  \ifx\combined@tempa\combined@first\else
  \ifx\combined@tempa\combined@last\combined@origlabel{SM:#1}%
  \else\combined@origlabel{#1}\fi\fi}
\let\combined@origref\ref
\def\ref#1{%
  \def\combined@tempa{#1}%
  \ifx\combined@tempa\combined@last\combined@origref{SM:#1}%
  \else\combined@origref{#1}\fi}
\let\combined@origmaketitle\maketitle
\let\@date\@empty                 
\def\normalsize{%
  \@setfontsize\normalsize\@xpt\@xiipt
  \abovedisplayskip 10\p@ \@plus2\p@ \@minus5\p@
  \belowdisplayskip \abovedisplayskip
  \abovedisplayshortskip  \abovedisplayskip
  \belowdisplayshortskip \abovedisplayskip
  \let\@listi\@listI
}\normalsize
\let\frontmatter@footnote@produce\frontmatter@footnote@produce@endnote
\def\bibsection{%
  \par
  \begingroup
    \baselineskip26\p@
    \bib@device{\hsize}{72\p@}%
  \endgroup
  \nobreak\@nobreaktrue
  \addvspace{19\p@}%
}
\@booleanfalse\twocolumn@sw
\set@page@grid
\def\maketitle{\setcounter{page}{1}\combined@origmaketitle\onecolumngrid\vspace{-18pt}}
\makeatother

\SMPrefixCitations

\title{Supplemental Material for\\ ``Interaction-driven ordering of multiple pseudospins and anomalous Hall metals in Bernal-stacked graphene''}

\author{Jeonghyeon Suh\,\orcidlink{0000-0001-8377-7402}}
\thanks{These authors contributed equally to this work.}
\affiliation{
Department of Physics and Astronomy,
\burl{https://ror.org/04h9pn542}
{Seoul National University},
Seoul 08826, Korea
}
\affiliation{
Research Institute of Basic Sciences,
\burl{https://ror.org/04h9pn542}
{Seoul National University},
Seoul 08826, Korea
}

\author{Kyungjin Shin\,\orcidlink{0009-0001-2368-0729}}
\thanks{These authors contributed equally to this work.}
\affiliation{
Department of Physics and Astronomy,
\burl{https://ror.org/04h9pn542}
{Seoul National University},
Seoul 08826, Korea
}

\author{Hongki Min\,\orcidlink{0000-0001-5043-2432}}
\email[Contact author: ]{hmin@snu.ac.kr}
\affiliation{
Department of Physics and Astronomy,
\burl{https://ror.org/04h9pn542}
{Seoul National University},
Seoul 08826, Korea
}

\maketitle

\section{Model and numerical details}
\label{app:model}

The effective tight-binding Hamiltonian in Eq.~(1) of the main text can be written explicitly in the layer--sublattice basis
$(1\mathrm{A},1\mathrm{B},\ldots,4\mathrm{A},4\mathrm{B})$ as
\begin{equation}
    \varepsilon_{\tau \bm{k} ; \sigma\sigma^{\prime}}^{(0)} =
    \begin{pmatrix}
        0 & v_0 \pi^\dagger & -v_4 \pi^\dagger & v_3 \pi & \gamma_2 / 2 & 0 & 0 & 0 \\
        v_0 \pi & \delta & \gamma_1 & -v_4 \pi^\dagger & 0 & \gamma_5 / 2 & 0 & 0 \\
        -v_4 \pi & \gamma_1 & \delta & v_0 \pi^\dagger & -\tilde{v}_4 \pi & \tilde{\gamma}_1 & \gamma_5 / 2 & 0 \\
        v_3 \pi^\dagger & -v_4 \pi & v_0 \pi & 0 & \tilde{v}_3 \pi^\dagger & -\tilde{v}_4 \pi & 0 & \gamma_2 / 2 \\
        \gamma_2 / 2 & 0 & -\tilde{v}_4 \pi^\dagger & \tilde{v}_3 \pi & 0 & v_0 \pi^\dagger & -v_4 \pi^\dagger & v_3 \pi \\
        0 & \gamma_5 / 2 & \tilde{\gamma}_1 & -\tilde{v}_4 \pi^\dagger & v_0 \pi & \delta & \gamma_1 & -v_4 \pi^\dagger \\
        0 & 0 & \gamma_5 / 2 & 0 & -v_4 \pi & \gamma_1 & \delta & v_0 \pi^\dagger \\
        0 & 0 & 0 & \gamma_2 / 2 & v_3 \pi^\dagger & -v_4 \pi & v_0 \pi & 0
    \end{pmatrix}_{\sigma\sigma^\prime},
\end{equation}
where $\pi \equiv \hbar (\tau k_x + i k_y)$ and
$v_i=(\sqrt{3}/2)a\gamma_i/\hbar$.
Here, $a=0.246~\mathrm{nm}$ is the graphene lattice constant, and $\tilde{v}_i$ is defined analogously in terms of $\tilde{\gamma}_i$.
The tight-binding parameters adopted from Ref.~\cite{Wu2015} are listed in Table~\ref{tab:TB_parameters}.
The parameters $\gamma_{1,3,4}$ correspond to the surface interfaces between layers 1--2 and 3--4, whereas $\tilde{\gamma}_{1,3,4}$ correspond to the bulk-like inner interface between layers 2--3.
For thicker Bernal multilayers, only the outermost interfaces, 1--2 and $(N-1)$--$N$, are treated as surface-like; all others are bulk-like, with $\gamma_{2,5}$ and the remaining parameters unchanged.
The perpendicular electric field adds an external interlayer potential to the diagonal of $\varepsilon^{(0)}$, with on-site energies ranging linearly from $-U_{\mathrm{ext}}$ on the bottom layer to $+U_{\mathrm{ext}}$ on the top layer, driving electrons toward the bottom layer and thus favoring upward layer polarization for $U_{\mathrm{ext}}>0$.
The Hartree--Fock equations are solved self-consistently at each $U_{\mathrm{ext}}$, which is also used to parameterize the field dependence of the results.
\begin{table}[htbp]
    \caption{Tight-binding parameters used for Bernal-stacked graphene in this work.}
    \label{tab:TB_parameters}
    \begin{ruledtabular}
    \renewcommand{\arraystretch}{1.3}
    \begin{tabular}{cccccccccc}
    $\gamma_0$ & $\gamma_1$ & $\gamma_2$ & $\gamma_3$ & $\gamma_4$ & $\gamma_5$ & $\delta$ & $\tilde{\gamma}_1$ & $\tilde{\gamma}_3$ & $\tilde{\gamma}_4$ \\ \colrule
    $3180~\mathrm{meV}$ & $390~\mathrm{meV}$ & $-12~\mathrm{meV}$ & $300~\mathrm{meV}$ & $40~\mathrm{meV}$ & $20~\mathrm{meV}$ & $50~\mathrm{meV}$ & $385~\mathrm{meV}$ & $250~\mathrm{meV}$ & $30~\mathrm{meV}$
    \end{tabular}
    \end{ruledtabular}
\end{table}
%

Electron--electron interactions are described by a layer-dependent Coulomb potential in a symmetric dual-gate geometry with two metallic gates at $z=\pm D_{\mathrm{g}}$,
where we take $D_{\mathrm{g}}=36.9~\mathrm{nm}$.
The interaction between layers $\ell$ and $\ell'$ takes the form~\cite{Kolar2023,Kwan2025b}
\begin{equation}
    \label{eq:Coulomb}
    V_{\ell\ell'}(q)
    =
    \frac{2\pi e^2}{\epsilon q}
    \left[
    e^{-q|z_{\ell}-z_{\ell'}|}
    +
    \frac{
    e^{-2qD_{\mathrm g}}
    \cosh\!\left(q(z_{\ell}-z_{\ell'})\right)
    -
    \cosh\!\left(q(z_{\ell}+z_{\ell'})\right)
    }{
    \sinh\!\left(2qD_{\mathrm g}\right)
    }
    \right].
\end{equation}
Here, $z_\ell=[\ell-(N+1)/2]d$ denotes the vertical coordinate of layer $\ell=1,\ldots,N$, and $d=0.335~\mathrm{nm}$ is the interlayer separation.
To reproduce the interaction-induced gap of approximately $5.2$~meV inferred from experiment~\cite{Nam2018}, we use an effective dielectric constant $\epsilon=9.6$ for the B4G calculations.
This interaction enters the Hartree and Fock terms in Eq.~(3) of the main text as $V_{\ell\ell'}(0)$ and $V_{\ell\ell'}(|\bm{k}-\bm{k}^{\prime}|)$, respectively.

We include intervalley interactions only in the LAF--QAH comparison in Fig.~4 of the main text, where we model them as valley-interchange (VI) exchange.
All other calculations, such as those for the phase diagram in Fig.~1, omit this term.
The large momentum transfer
$|\bm{K}-\bm{K}'|=4\pi/(3a)\gg|\bm{k}-\bm{k}'|$
allows us to treat this exchange as short-ranged~\cite{Yang2025,Dutta2026}.
We parameterize its strength as
$V_{\mathrm{VI}}=\alpha_{\mathrm{VI}}\,V_{\ell\ell}(|\bm{K}-\bm{K}'|)$,
where $\alpha_{\mathrm{VI}}$ accounts for lattice-scale contributions
beyond the continuum Coulomb interaction.
At this momentum transfer, interlayer components are suppressed by $e^{-|\bm{K}-\bm{K}'|d}\approx3\times10^{-3}$ and are therefore neglected.
Gate screening is also negligible, giving $V_{\mathrm{VI}}\simeq2\pi e^{2}\alpha_{\mathrm{VI}}/(\epsilon|\bm{K}-\bm{K}'|)$.
The resulting VI exchange term in the mean-field Hamiltonian is
\begin{equation}
    \varepsilon^{(\mathrm{VI})}_{\tau s;\sigma\sigma'}
    = -V_{\mathrm{VI}}\,\delta_{\ell\ell'}\sum_{\bm{k}}
    \left<\hat{c}^{\dagger}_{-\tau s\bm{k};\sigma'}\hat{c}_{-\tau s\bm{k};\sigma}\right> ,
\end{equation}
where $\ell$ ($\ell'$) is the layer index of $\sigma$ ($\sigma'$).

The self-consistent solutions are obtained by iteratively diagonalizing $\hat{H}_{\mathrm{MF}}$ and updating the density matrix on a $241\times241$ $\bm{k}$-point mesh centered at each valley,
with the optimal damping algorithm employed to stabilize convergence~\cite{Cances2000}.
We identify the ground state as the lowest-energy solution among those obtained from different initial states.
Representative total-energy comparisons between competing flavor orders are presented in Sec.~\ref{sec:Etot_compare}.

\section{Pseudospin ordering and energetics}
\label{sec:Etot_compare}

\begin{figure}[t!]
    \centering
    \includegraphics[width=0.95\linewidth]{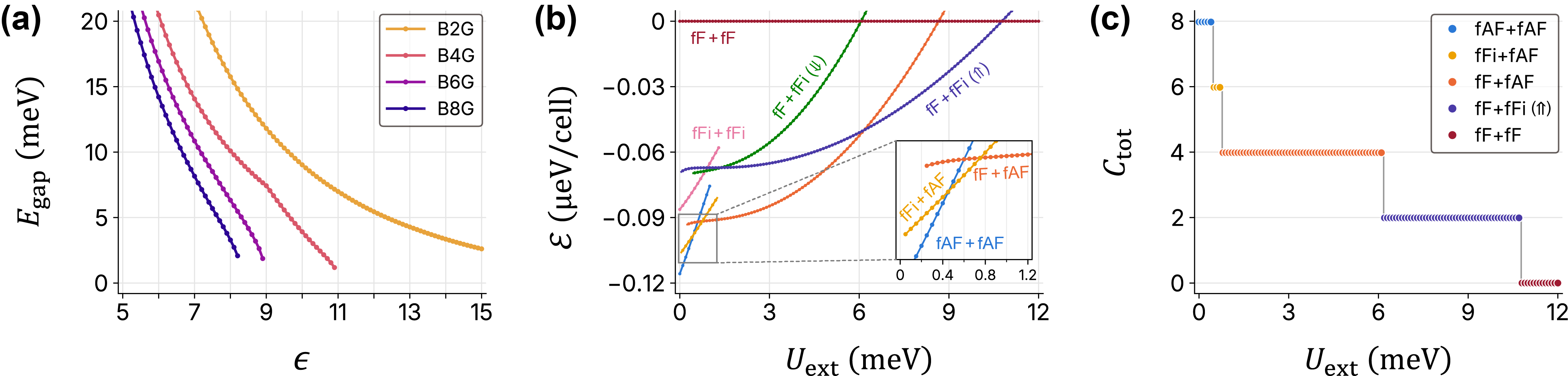}
    \caption{
    (a) Interaction-induced band gap $E_{\mathrm{gap}}$ of B2G, B4G, B6G, and B8G at the charge neutrality point (CNP) and $U_{\mathrm{ext}}=0$ as a function of the dielectric constant $\epsilon$.
    All curves except that for B2G terminate just before the gap closes, beyond which the lightest pseudospin sector undergoes a band inversion.
    (b) Total energy per unit cell $\mathcal{E}$ of competing flavor orders in B4G at the CNP relative to the ``fF+fF'' state, as a function of $U_{\mathrm{ext}}$.
    The inset resolves the energy crossings that give the fAF--fFi--fF sequence of the light-mass doublet.
    Two ``fF+fFi'' configurations differ in the layer polarization of the majority heavy-mass flavors, indicated by arrows.
    (c) Ground-state Chern number $C_{\mathrm{tot}}$ along the ground-state ordering sequence in (b), starting from the QAH state at $U_{\mathrm{ext}}=0$.
    Colors in (b) and (c) denote the flavor orders, with additional metastable orders shown in green and pink in (b).
    }
    \label{figS1}
\end{figure}

Figure~\ref{figS1}(a) shows the interaction-induced band gap $E_{\mathrm{gap}}$ of even-layer Bernal multilayers at the CNP and $U_{\mathrm{ext}}=0$ as a function of the dielectric constant $\epsilon$.
The gap decreases with increasing $\epsilon$, reflecting the weakening of the exchange interaction that drives pseudospin ordering, and also decreases with increasing layer number at fixed $\epsilon$.
For B4G, B6G, and B8G, the gap persists up to $\epsilon\approx10.9$, $8.9$, and $8.2$, respectively,
beyond which it closes and the lightest pseudospin sector undergoes a band inversion, so that the multi-doublet picture no longer applies even at the CNP.
In contrast, B2G remains gapped well beyond the plotted range, up to at least $\epsilon=25$.
The parameter range supporting this picture thus narrows with increasing layer number, as the interaction-induced gaps become insufficient to overcome remote-hopping effects.

Our choice of $\epsilon=9.6$ for B4G lies below the critical value $\epsilon\approx10.9$,
but the resulting gap is comparable to $|\gamma_{2}|/\sqrt{5}\approx5.4$~meV, the magnitude of the diagonal shift of each doublet due to $\gamma_2$.
As a result, the heavy-mass conduction band lies close to the light-mass one, so that only the hole-doped side remains within the interaction-dominant regime for the parameters considered.

Figure~\ref{figS1}(b) compares the total energies of competing self-consistent flavor orders in B4G at the CNP as a function of $U_{\mathrm{ext}}$, using the ``fF+fF'' state as the energy reference.
Here, the first and second labels denote the orders of the light- and heavy-mass doublets, respectively.
At $U_{\mathrm{ext}}=0$, the ``fAF+fAF'' state is the ground state, in which the two doublets are polarized toward opposite layers within each flavor owing to inter-doublet exchange (see Sec.~\ref{sec:doublet_AF}).
As $U_{\mathrm{ext}}$ increases, the light-mass doublet first undergoes the fAF--fFi--fF sequence at $U_{\mathrm{ext}}\approx0.5$ and $0.8$~meV (inset), yielding the ``fFi+fAF'' and ``fF+fAF'' states while the heavy-mass doublet retains fAF order.
The heavy-mass doublet then undergoes the same sequence at $U_{\mathrm{ext}}\approx6.1$ and $10.8$~meV, yielding the ``fF+fFi'' and ``fF+fF'' states.
The order-of-magnitude difference between the critical potentials of the two doublets reflects the larger weight of the light-mass doublet on the outer layers, which couple most strongly to $U_{\mathrm{ext}}$.

For the ordering sequence starting from the QAH state at $U_{\mathrm{ext}}=0$, each transition reverses the layer polarization of one pseudospin sector in a single flavor, flipping its Chern contribution and reducing $C_{\mathrm{tot}}$ by 2,
so that $C_{\mathrm{tot}}$ decreases stepwise as $8\rightarrow6\rightarrow4\rightarrow2\rightarrow0$, as shown in Fig.~\ref{figS1}(c).
The ``fFi+fAF'' state with $C_{\mathrm{tot}}=6$ [region (a) in Fig.~1 of the main text] thus occupies only the narrow interval $0.5\lesssim U_{\mathrm{ext}}\lesssim0.8$~meV between the two light-mass transitions.

\section{Inter-doublet ordering}
\label{sec:doublet_AF}

To elucidate the inter-doublet ordering, we consider the minimal model retaining only $\gamma_{0}$ and $\gamma_{1}$,
in which the low-energy B4G Hamiltonian decouples exactly into two bilayer-like pseudospin sectors~\cite{Min2008a,Min2008b}.
The low-energy states of the light- and heavy-mass sectors reside on the following superpositions of the nondimer orbitals:
\begin{equation}
\label{eq:B4G_pseudospin_basis}
\begin{aligned}
    |\mathrm{A}^{\mathrm L}\rangle &= \eta_1 |1\mathrm{A}\rangle - \eta_2 |3\mathrm{A}\rangle,
    &\qquad
    |\mathrm{B}^{\mathrm L}\rangle &= \eta_2 |2\mathrm{B}\rangle - \eta_1 |4\mathrm{B}\rangle,
    \\
    |\mathrm{A}^{\mathrm H}\rangle &= \eta_2 |1\mathrm{A}\rangle + \eta_1 |3\mathrm{A}\rangle,
    &\qquad
    |\mathrm{B}^{\mathrm H}\rangle &= \eta_1 |2\mathrm{B}\rangle + \eta_2 |4\mathrm{B}\rangle ,
\end{aligned}
\end{equation}
with $|\eta_{1}|^{2} = (5+\sqrt{5})/10 \simeq 0.724$ and $|\eta_{2}|^{2} = (5-\sqrt{5})/10 \simeq 0.276$.
Since $|\eta_{1}|>|\eta_{2}|$, the light- and heavy-mass doublets have greater weight on the outer and inner nondimer orbitals, respectively.
Among the remote hoppings, only $\gamma_{2}$ couples the two sectors appreciably near the band edges, and it does so only within the same sublattice
($\mathrm{A}^{\mathrm L}\leftrightarrow\mathrm{A}^{\mathrm H}$, $\mathrm{B}^{\mathrm L}\leftrightarrow\mathrm{B}^{\mathrm H}$).
Couplings between different sublattices of the two sectors arise only at finite $\bm{k}$ and are negligible,
so that the sublattice structure underlying the argument below is preserved, consistent with the inter-doublet ordering found in the full-model HF calculations (Fig.~2 of the main text).

The inter-doublet ordering originates from the interplay between the layer-resolved overlap of the light- and heavy-mass pseudospin wave functions and the layer-dependent Coulomb interaction.
To see this directly, we project the Fock exchange onto the two pseudospin sectors.
For a fixed spin--valley flavor, the diagonal pseudospin occupations contribute to the inter-doublet Fock energy as
\begin{equation}
\label{eq:EF_LH}
    E_{\mathrm F}^{\mathrm{LH}}
    =
    -\sum_{\bm{k},\bm{k}',\alpha,\beta}
    \mathcal{V}_{\alpha\beta}^{\mathrm{LH}}
    (|\bm{k}-\bm{k}'|)
    n_{\alpha}^{\mathrm L}(\bm{k})
    n_{\beta}^{\mathrm H}(\bm{k}'),
\end{equation}
where $n_{\alpha}^{r}(\bm{k})$ denotes the occupation of $|\alpha^{r}\rangle$, with $\alpha=\mathrm{A},\mathrm{B}$ in doublet $r=\mathrm{L},\mathrm{H}$.
The exchange kernel in the $\mathrm{AA}$ channel is
\begin{align}
\label{eq:VAA_LH}
    \mathcal{V}_{\mathrm{AA}}^{\mathrm{LH}}(q)
    &=
    \sum_{\ell,\ell'}
    \bigl(u_{\ell,\mathrm A}^{\mathrm L}\bigr)^*
    u_{\ell,\mathrm A}^{\mathrm H}
    V_{\ell\ell'}(q)
    \bigl(u_{\ell',\mathrm A}^{\mathrm H}\bigr)^*
    u_{\ell',\mathrm A}^{\mathrm L}
    \nonumber\\
    &=
    \eta_1^2\eta_2^2
    \left[
        V_{11}(q)
        -
        V_{13}(q)
        -
        V_{31}(q)
        +
        V_{33}(q)
    \right]
    \nonumber\\
    &\simeq
    \frac{2}{5}V_{\mathrm{s}}(q)
    \left(1-e^{-2qd}\right)
    >0 .
\end{align}
Here $u_{\ell,\alpha}^{r}=\langle \ell\alpha|\alpha^{r}\rangle$.
In the last line, $V_{\mathrm{s}}(q)$ denotes the same-layer interaction, with the layer dependence approximated as
$V_{\ell\ell^{\prime}}(q)\simeq V_{\mathrm{s}}(q)e^{-q|\ell-\ell^{\prime}|d}$.
The $\mathrm{BB}$ channel gives the same kernel,
$\mathcal{V}_{\mathrm{BB}}^{\mathrm{LH}}(q)
=
\mathcal{V}_{\mathrm{AA}}^{\mathrm{LH}}(q)$.
By contrast, the mixed-sublattice kernels vanish,
$\mathcal{V}_{\mathrm{AB}}^{\mathrm{LH}}(q)
=
\mathcal{V}_{\mathrm{BA}}^{\mathrm{LH}}(q)
=
0$,
reflecting the absence of common layer support.

Because the Fock term enters with a negative sign, the finite same-sublattice exchange favors a common sublattice polarization of the light- and heavy-mass doublets.
This common sublattice polarization, however, corresponds to opposite layer polarizations of the two doublets due to their distinct layer compositions.
More generally, the same pseudospin structure extends to thicker even-layer Bernal graphene; in B6G, for example, the A- and B-sublattice components of the three bilayer-like doublets are distinct superpositions of the $(1\mathrm{A},3\mathrm{A},5\mathrm{A})$ and $(2\mathrm{B},4\mathrm{B},6\mathrm{B})$ nondimer orbitals, respectively.
The inter-doublet exchange kernels remain positive in the same-sublattice channels and vanish in the mixed-sublattice channels, favoring a common sublattice polarization among the doublets.
Consequently, the layer polarization alternates among the bilayer-like pseudospin doublets in order of effective mass, giving rise to the characteristic inter-doublet ordering of even-layer Bernal multilayers, as reported in Ref.~\cite{Yoon2017}.


\end{document}